\documentclass[sigconf, authorversion=true]{acmart}
   
\AtBeginDocument{%
  \providecommand\BibTeX{{%
    \normalfont B\kern-0.5em{\scshape i\kern-0.25em b}\kern-0.8em\TeX}}}
    
\copyrightyear{2021} 
\acmYear{2021} 
\setcopyright{acmlicensed}\acmConference[ICPE '21 Companion]{Companion of the 2021 ACM/SPEC International Conference on Performance Engineering}{April 19--23, 2021}{Virtual Event, France}
\acmBooktitle{Companion of the 2021 ACM/SPEC International Conference on Performance Engineering (ICPE '21 Companion), April 19--23, 2021, Virtual Event, France}
\acmPrice{15.00}
\acmDOI{10.1145/3447545.3451189}
\acmISBN{978-1-4503-8331-8/21/04}

\usepackage{amsmath}

\usepackage{amssymb,amsfonts}

\usepackage{algorithmic}

\usepackage{graphicx}

\usepackage{textcomp}
\usepackage{xcolor}
\usepackage{float}
\usepackage{tikz}
\usepgflibrary{arrows}

\usetikzlibrary{arrows.meta}
\usepackage{todonotes}
\usepackage{listings}

\usepackage[latin1]{inputenc}
\usepackage{tikz}
\usetikzlibrary{shapes,arrows}

\title[PIERES]{PIERES: A Playground for Network Interrupt Experiments on Real-Time Embedded Systems in the IoT}

\author{Franz Bender}
\email{franz.bender@icloud.com}

\affiliation{%
  \institution{Technische Universit\"at Berlin}
  \city{Berlin}
  \country{Germany}}

\author{Jan Jonas Brune}
\email{jan.j.brune@campus.tu-berlin.de}

\affiliation{%
  \institution{Technische Universit\"at Berlin}
  \city{Berlin}
  \country{Germany}}

\author{Nick Lauritz Keutel}
\email{keutel@campus.tu-berlin.de}

\affiliation{%
  \institution{Technische Universit\"at Berlin}
  \city{Berlin}
  \country{Germany}}

\author{Ilja Behnke}
\email{i.behnke@tu-berlin.de}

\affiliation{%
  \institution{Technische Universit\"at Berlin}
  \city{Berlin}
  \country{Germany}}

\author{Lauritz Thamsen}
\email{lauritz.thamsen@tu-berlin.de}

\affiliation{%
  \institution{Technische Universit\"at Berlin}
  \city{Berlin}
  \country{Germany}}

\begin{document}
\begin{abstract}
IoT devices have become an integral part of our lives and the industry. Many of these devices run real-time systems or are used as part of them. As these devices receive network packets over IP networks, the network interface informs the CPU about their arrival using interrupts that might preempt critical processes. Therefore, the question arises whether network interrupts pose a threat to the real-timeness of these devices. However, there are few tools to investigate this issue.

We present a playground which enables researchers to conduct experiments in the context of network interrupt simulation. The playground comprises different network interface controller implementations, load generators and timing utilities. It forms a flexible and easy to use foundation for future network interrupt research. We conduct two verification experiments and two real world examples. The latter give insight into the impact of the interrupt handling strategy parameters and the influence of different load types on the execution time with respect to these parameters.
\end{abstract}



\maketitle
\pagestyle{plain}

\section{Introduction}

Many processes in industrial settings have real-time constraints. Engineers need guarantees for these processes, e.g. that an action $A$ takes at most $N$ seconds. Real-time operating systems (RTOSs) have proven to be a suitable platform for implementing software by providing such guarantees. RTOSs offer, in contrast to regular operating systems, special interfaces for precise timing and scheduling of time critical tasks~\cite{stankovicRealTimeOperatingSystems}. These RTOSs have therefore found application in the field of embedded systems such as sensor systems, industrial control systems and many other specialized devices. For researchers and engineers using RTOSs it is essential that the real-timeness of their device is maintained.

With the rise of the Internet of Things (IoT) many of these devices get connected to the internet. This trend has been described as a new era~\cite{wollschlaegerFutureIndustrialCommunication2017}. It enables features such as remote monitoring, remote debugging and a more intelligent management of devices by combining data from multiple devices for decisions. You could take smart navigation systems as proposed in \cite{handteInternetofThingsEnabledConnected2016} as an example for the latter. 

Modern network interface controllers (NIC) such as the Intel\copyright 82574 GbE Controller Family use interrupts to inform the CPU about new packets. This implies that other network devices have the ability to invoke interrupts at the target. As interrupts are handled by interrupt service routines (ISRs), which have a very high execution priority, these interrupts may interfere with user code. As the rate of network interrupts increases, more time is spent in the ISR instead of the user code~\cite{behnkeInterruptingRealTimeIoT}.

This behavior is sound as ISRs reside in a higher priority space to minimize the I/O delay of the embedded system. However, while being a valid phenomenon, this behavior may not be intended by the engineer of the device, as network packets might not be as important as the critical user code. Note that the high number of interrupts may be caused by a malicious attacker or by other non-malicious conditions such as a bad network configuration or other similar conditions. All in all it means that adding network capabilities to real-time systems adds a per-packet workload which may drown critical tasks and break time guarantees.

In this paper we present a playground which enables researchers and engineers to tackle these issues. The playground
\begin{itemize}
	\item can simulate network interrupts on a microcontroller to help analyze the impact of network traffic on applications running on IoT devices.
	\item offers simulation of continuous and Poisson-distributed\footnote{A Poisson distribution is commonly used to model traffic in the literature, e.g. \cite{rajaratnamNetworkModellingCircuitswitched1996}} network interrupts, as well as replays of captures of actual network traffic
\end{itemize}

The remainder of this paper is structured as follows. Section II gives an overview of the related work%
Section III details the approach we took to design the playground. Section IV details the experiments we conducted using our playground. Finally, Section V summarizes our work.%

\section{Related Work}

Available literature focusses mostly on mitigation strategies.

As a general example, \textit{interrupt moderation} or \textit{coalescing} \cite{salahCoalesceNotCoalesce2007} is used by the authors of \cite{hanIntervalbasedAdaptiveInterrupt2016} to reduce energy consumption in a system for high speed networks. Multiple interrupts are grouped to invoke a single interrupt at a later time.

Other approaches rely on the optimization of the interrupts themselves. The authors of \cite{nakashimaDesignImplementationInterrupt2002} use that the interrupt mechanism can be split up into preprocessing, the ISR and postprocessing. They exploit that the preprocessing and the postprocessing are independent of the interrupt itself and ``reuse'' these phases for multiple interrupts and only calling the ISR for every interrupt. 

While these authors worked on mitigation strategies, others conducted research on the interrupts themselves. This resulted in different models for network interrupts that differ in their approaches: The authors of \cite{houAnalysisInterruptBehavior2018} present a complex model for the firing and execution of interrupts. They use an extension of stochastic Petri nets to model the system as this allows them to combine the probabilistic aspects caused by the randomness of the interrupts with the stateful aspects of prioritized interrupt handling. 

A more basic model was presented in \cite{rajaratnamNetworkModellingCircuitswitched1996}. There, a double interrupted poisson process is employed for their calculations. This is a poisson process that can be can be in either a high or a low state which determines the poisson parameters for the load. They conclude by giving formulas for blocking probabilities, i.e. the probability that a link is blocked and packets are dropped.

To the best of our knowledge, there is no work that focusses on providing a testing capability for the impact of network interrupts on real-time systems.

\section{Approach}

This section describes the goal of the playground, how to enable different NIC implementations and the choice of network traffic scenarios as test loads, which are detailed in the following.

\subsection{Goal of the Playground}
The goal of this playground is to enable researchers to conduct experiments and to enable engineers to test their code regarding network interrupt simulation. This shall be done on an actual device used for IoT applications. Therefore, the playground has to fulfill the following requirements: It has to
\begin{itemize}
	\item run on real IoT hardware.
	\item be capable of simulating multiple NIC implementations.
	\item be able to simulate multiple network traffic scenarios.
	\item be easily configurable.
	\item have minimal performance impact on the tested process.
\end{itemize}

\begin{figure}[t]
	\begin{center}
	\tikzstyle{block} = [rectangle, draw, fill=black!10, 
    text width=10em, text centered, rounded corners, minimum height=1.5em]
	\tikzstyle{line} = [draw, -latex']
    
	\begin{tikzpicture}[node distance = 1.5cm, auto, every node/.style={font=\normalsize}]
		
	    \node [block] (setup) {setup microcontroller};
    	\node [block, below = 1em of setup] (insert) {insert user code};
	    \node [block, below = 1em of insert] (randLoad) {configure load};
	    \node [block, right = 1em of randLoad, node distance=3cm] (pcap) {parse recorded PCAP};
    	\node [block, below = 1em of randLoad] (nic) {configure NIC};
    	\node [block, below = 1em of nic] (flash) {flash to \textmu{}C \& execute};
    	\node [block, below = 1em of flash] (analysis) {copy result for analysis};
    	\path [line] (setup) -- (insert);
	    \path [line] (insert) -- (randLoad);
    	\path [line] (randLoad) -- (nic);
	    \path [line] (pcap) |- (nic);
    	\path [line] (insert) -| (pcap);
	    \path [line] (nic) -- (flash);
	    \path [line] (flash) -- (analysis);
	\end{tikzpicture}
	\end{center}
	\caption{General procedure of a user operating the playground.}
	\label{process}
\end{figure}
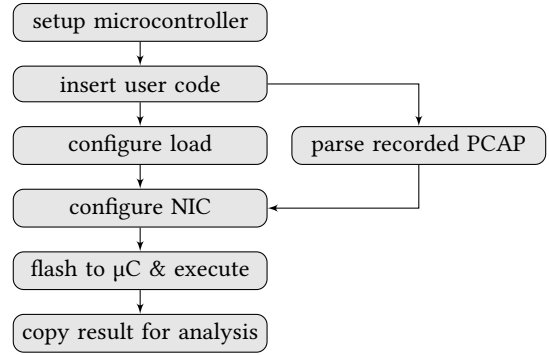

\subsection{Operation of the Playground}
The operation of the playground is shown in Fig. \ref{process}. The playground is setup on the microcontroller. The user then inserts the code of the critical process under test and either chooses a configuration for a Uniform, a Random Load, or parses a recorded network trace. Afterwards, the user configures the NIC and flashes the playground onto the device. Measured metrics can be configured by the user and range from execution time and number of interrupts to more complex metrics such as the ratio between interrupt sources (see Sec.\ref{sec:experiments}).

\subsection{NIC Implementations}
The playground has to model different NICs, that can be configured by the user. The simplest NIC would inform the CPU at the arrival of every packet by triggering an interrupt for each. Alternatively, some modern NICs offer interrupt moderation. To incorporate this into the playground, we offer the choice between a simple NIC model without any interrupt moderation and several smarter NIC implementations.

For the simple NIC, the duration $d(l)$ of an interrupt is dependent on the packet length $l$ and plainly modeled as a length dependent delay $d_l$ which is evoked $l$ times plus constant length independent delay $d_c$ which is an overhead evoked for every interrupt:

\begin{equation*}
	d(l) = d_l \cdot l + d_c
\end{equation*}

Different simple NIC implementations can be characterized by the user through setting the values for the length dependent delay $d_l$ and the length independent delay $d_c$. %

For smarter, more elaborate NIC simulations with interrupt moderation, we support NICs to be defined with a counter mode, a timer mode or a combination out of the box. Here, the parts of the interrupt duration model, packet length and corresponding dependent and independent delay, are modeled twice each, once for the simulated ISR and once for a simulated receiver task.

A NIC with the counter mode does not trigger an interrupt for every received packet, but counts the arriving packets, stores them in a buffer and -- after a specified number of packets -- evokes one interrupt for them all. Afterwards, the counter and buffer are reset. Another option is a NIC with the timer mode. In this case, for an arriving packet a delay timer of specified duration is set. Upon expiration, one interrupt will be triggered. If further packets arrive before the timer has run out, the timer will be reset without evoking an interrupt. However, problems may arise if the timer is constantly being reset by arriving packets, never allowing an interrupt to be triggered. The combination of both modes circumvents this problem. %

\subsection{Network Traffic Scenarios as Load}
The arrival of packets with corresponding time stamps over some observed time constitutes a load scenario. As we want to simulate different scenarios, the playground offers uniform loads, random loads and user defined/recorded loads.

Uniform loads have a constant receive frequency. For the randomized loads, a Poisson distribution is used to model the arrival of new packets.

The Poisson distribution is achieved by inverse transform sampling with a uniform distribution. Assuming the number of incoming packets per interval $p_i$ are Poisson distributed, the inter-arrival time $d_i$ is exponentially distributed:

\begin{equation*}
\begin{split}
	p_i &\sim Poisson(\lambda), \; d_i = p_{i+1} - p_i\\
	\Rightarrow d_i &\sim Exp(\lambda)\\
	\hat{d}_i &= F^{-1}(u_i) = - \frac{1}{\lambda}\ln(1-u_i) \hat{=} -\frac{1}{\lambda}\ln(u_i).
\end{split}
\end{equation*}

By inverse transform sampling (as described in \cite{Murphy2012} Section 23.2) we determine the empirical delays between packets $\hat{d}_i$ by sampling $u_i \sim U(0,1)$ and calculating $\hat{d}_i$. By setting the parameter $\lambda$, different randomized Poisson distributed loads can be specified.

As a third option, the playground allows for recorded network scenarios to be replayed.

\section{Experiments}
\label{sec:experiments}

Two validation and two demonstration experiments were performed for which a large summation (in a loop) with a conditional statement at each step of the iteration was used as the user code.

\subsection{Prototype Implementation}
The playground is implemented on the ESP32 \footnote{ESP32 Series Datasheet Version 3.4, \url{https://www.espressif.com/sites/default/files/documentation/esp32_datasheet_en.pdf}, accessed 2021-01-07}, a dual core CPU, and written in C and C++, matching the system of the microcontroller. The parsing script for the recorded network scenarios (PCAPs) is written in Python and generates C++ code. The last playground requirement is fulfilled by using both cores to separate the computational load of the playground code from the tested user code.

\subsection{Validation}

All validation testing was performed using a Poisson-distributed random load. The following results merely show a selection of all the experiments that have been carried out with the playground. 

\begin{figure}[t]
	\centering
	\includegraphics[width=0.5\textwidth, height=0.265\textwidth]{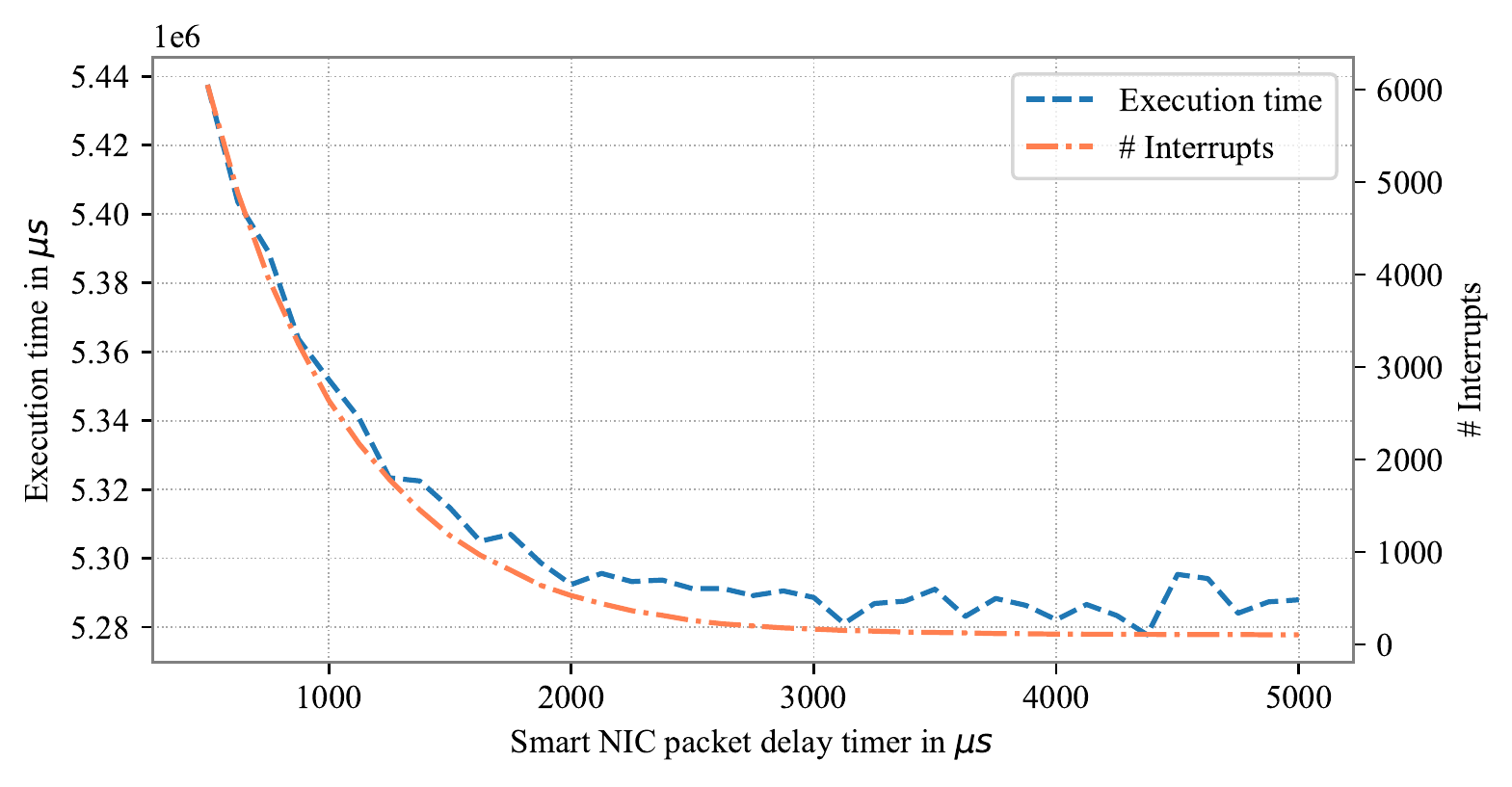}
	\caption{Execution time and number of interrupts decrease inversely with the increase of the times threshold when using the Smart NIC packet delay timer for interrupt moderation.}
	\label{packetTimeoutVsExcTimeAndIntrs}
\end{figure}

First, the impact of the packet delay timer on the execution time and the number of interrupts was measured, as seen in Fig. \ref{packetTimeoutVsExcTimeAndIntrs}. This test was performed using the combined interrupt moderation mode. %
Both, the execution time and the number of interrupts decrease with an increasing packet delay timer. As the number of packets is constant, the execution time is reduced by using interrupt moderation. This gain from increasing the packet delay timer comes with the caveat of a higher packet latency. %

\begin{figure}[t]
	\centering
	\includegraphics[width=0.5\textwidth, height=0.265\textwidth]{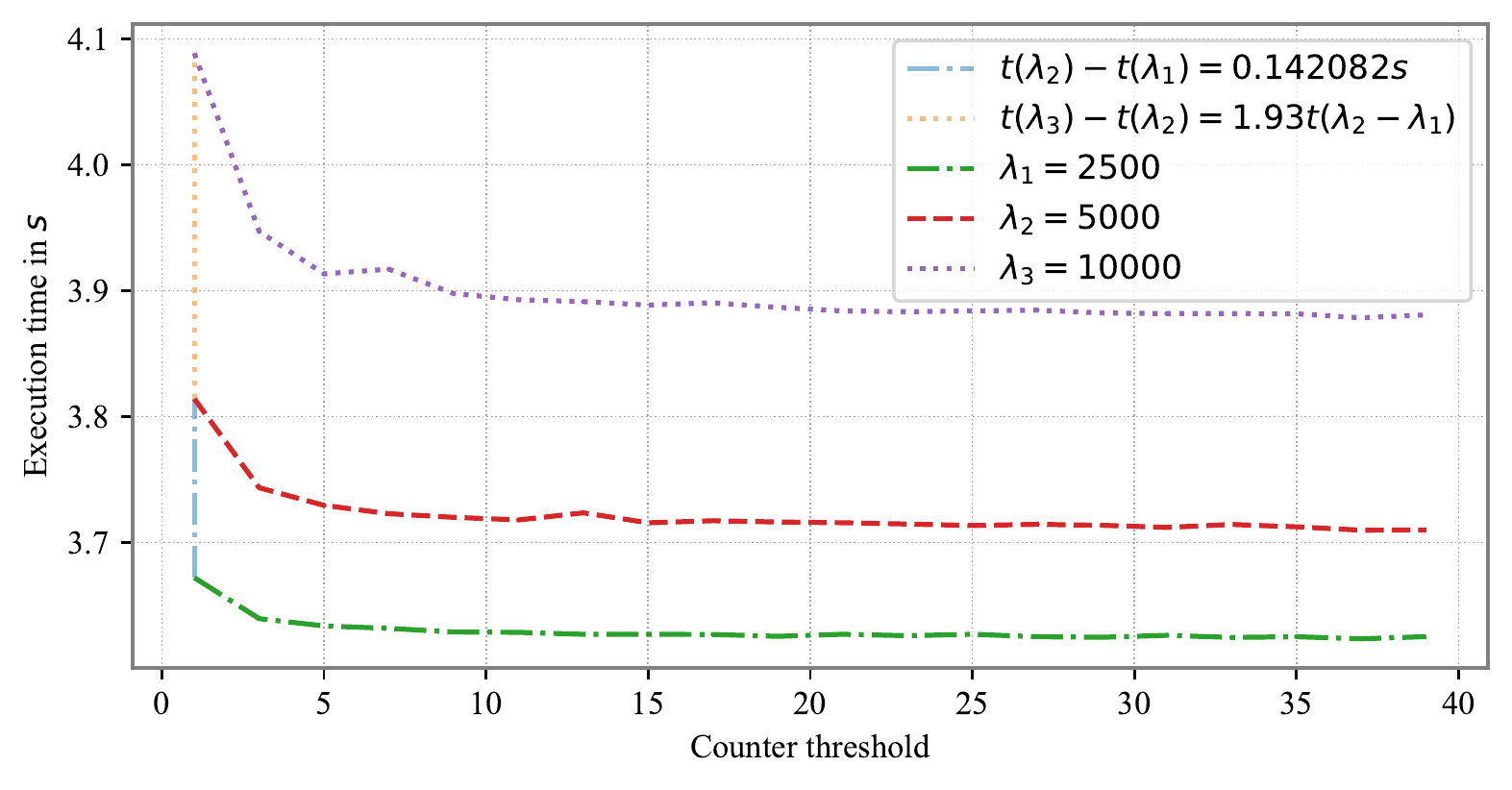}
	\caption{Comparison of the execution time of three Poisson-distributed random network load scenarios for different $\lambda$ in dependence of the counter threshold.}
	\label{counterRatioPreserv}
\end{figure}

In the second validation experiment three different parameters $\lambda$ for the Poisson-distributed random load generation were compared, as seen in Fig. \ref{counterRatioPreserv}. This test was performed using the counter mode. A larger parameter value corresponds to a higher load. In the graph, the counter threshold is plotted against the execution time. %
$\lambda_1$ is half as big as $\lambda_2$ and $\lambda_3$ is twice as big as $\lambda_2$. As shown in the legend, it can be observed that the difference between the execution times of $\lambda_3$ and $\lambda_2$ is twice as big as the one between $\lambda_2$ and $\lambda_1$. This ratio stays roughly the same along the x-axis. This shows that the counter mode scales linearly with load.

\subsection{Practical Examples}

\begin{figure}[t]
	\centering
	\includegraphics[width=0.5\textwidth, height=0.265\textwidth]{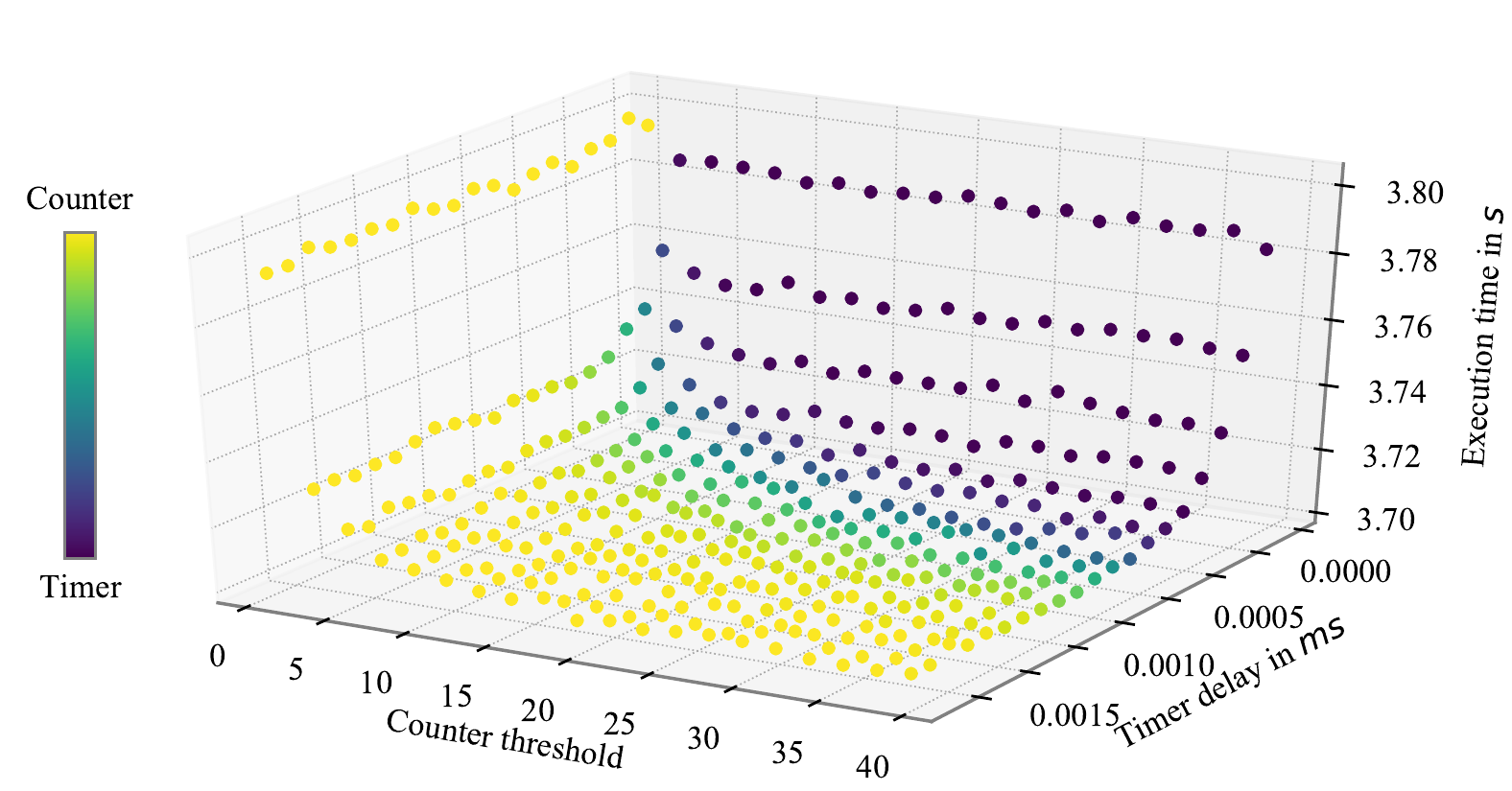}
	\caption{Ratio of the reasons for triggering interrupts, shown in relation to counter threshold, timer delay and execution time when using the smart NIC with mixed mode.}
	\label{prac2}
\end{figure}

In the first practical example the reasons for interrupts using the combined interrupt moderation mode of the smart NIC model were investigated, as seen in Fig. \ref{prac2}. A Poisson-distributed random load was used for this test. The coloring of the data points indicates the reason why an interrupt was triggered. An area in the plane of the data points where the coloring indicates an equilibrium between the two reasons can be observed. The area extends in both counter threshold and timer delay directions but drifts towards the direction of the counter threshold axis, indicating that with increasing counter threshold, it plays less of a role in causing interrupts than the timer delay does.%

\begin{figure}[t]
	\centering
	\includegraphics[width=0.5\textwidth, height=0.265\textwidth]{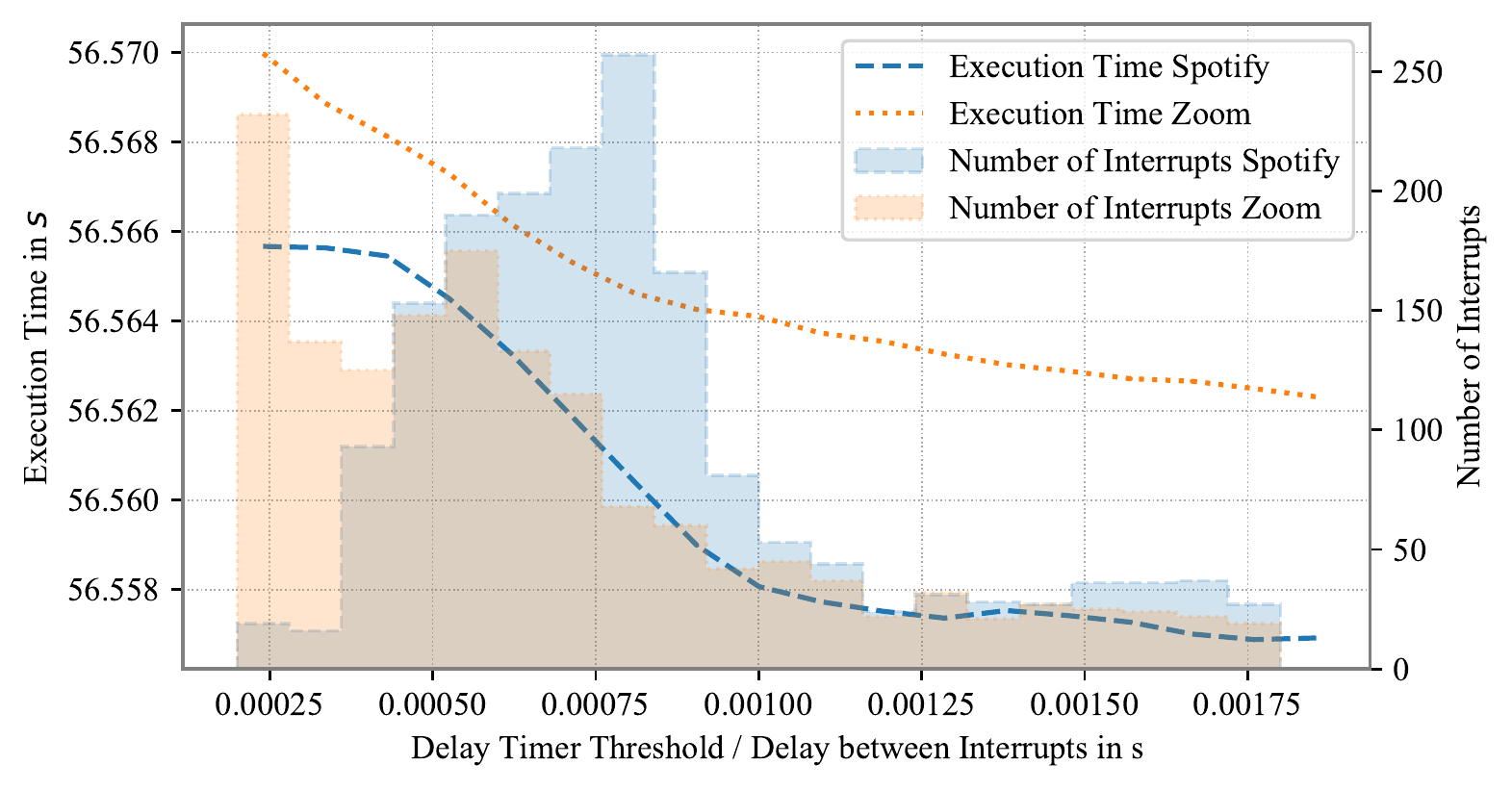}
	\caption{Execution time of the test code in two replayed network scenarios from prerecorded PCAPs is shown for diverse delay timer thresholds.}
	\label{pcapDiag}
\end{figure}

In the second practical example we take a look at recorded loads. We compare the execution time of the user code when using a mixed mode NIC with a Spotify network load to a Zoom conference load that have been prerecorded. The load is a lot less intense compared to the previous experiments. We use these two loads because they have two different packet arrival patterns: the Spotify load is a bursty load, while the Zoom load is a rather continuous load. Note that we use a longer running user code  (more iterations) here to allow for a longer measurement. 

Fig. \ref{pcapDiag} is a combination of two diagrams: the lines show the execution time for different delay timer thresholds while the filled area is a histogram which shows in what intervals packets arrive in the two load scenarios. We see that the execution time in the Spotify scenario benefits more from the relaxing of the timer timeout delay while in the Zoom scenario the behavior matches the behavior of a Poisson load more closely.

When comparing the results from both scenarios it becomes more obvious that the expected load progression can be used to find suitable interrupt moderation parameters.
\footnote{
	code and details on \href{https://github.com/dos-group/pieres_playground}{https://github.com/dos-group/pieres\_playground}
}

\section{Conclusion and Future Work}

We presented a playground which enables researchers to conduct experiments in the context of network interrupt simulation in real-time scenarios. It offers multiple load generators including random and custom prerecorded settings as well as logging capabilities. The playground was validated through a series of tests. We also presented two practical use cases, highlighting the ability of the playground to simulate desired network characteristics and analyze the results.

Further steps include a broader range of more complex NIC models and random load sources. Additionally, a repository of PCAP files could be created by playground users.

\bibliographystyle{ACM-Reference-Format}
\bibliography{9-bibliography/bibliography}


\begin{thebibliography}{10}


\ifx \showCODEN    \undefined \def \showCODEN     #1{\unskip}     \fi
\ifx \showDOI      \undefined \def \showDOI       #1{#1}\fi
\ifx \showISBNx    \undefined \def \showISBNx     #1{\unskip}     \fi
\ifx \showISBNxiii \undefined \def \showISBNxiii  #1{\unskip}     \fi
\ifx \showISSN     \undefined \def \showISSN      #1{\unskip}     \fi
\ifx \showLCCN     \undefined \def \showLCCN      #1{\unskip}     \fi
\ifx \shownote     \undefined \def \shownote      #1{#1}          \fi
\ifx \showarticletitle \undefined \def \showarticletitle #1{#1}   \fi
\ifx \showURL      \undefined \def \showURL       {\relax}        \fi
\providecommand\bibfield[2]{#2}
\providecommand\bibinfo[2]{#2}
\providecommand\natexlab[1]{#1}
\providecommand\showeprint[2][]{arXiv:#2}

\bibitem[\protect\citeauthoryear{Behnke, Pirl, Thamsen, Danicki, Polze, and
  Kao}{Behnke et~al\mbox{.}}{2020}]%
        {behnkeInterruptingRealTimeIoT}
\bibfield{author}{\bibinfo{person}{Ilja Behnke}, \bibinfo{person}{Lukas Pirl},
  \bibinfo{person}{Lauritz Thamsen}, \bibinfo{person}{Robert Danicki},
  \bibinfo{person}{Andreas Polze}, {and} \bibinfo{person}{Odej Kao}.}
  \bibinfo{year}{2020}\natexlab{}.
\newblock \showarticletitle{Interrupting {{Real}}-{{Time IoT Tasks}}: {{How Bad
  Can It Be}} to {{Connect Your Critical Embedded System}} to the
  {{Internet}}?}. In \bibinfo{booktitle}{\emph{IPCCC 2020: 39th International
  Performance Computing and Communications Conference}}. IEEE,
  \bibinfo{pages}{1--6}.
\newblock


\bibitem[\protect\citeauthoryear{Han and Kim}{Han and Kim}{6 10}]%
        {hanIntervalbasedAdaptiveInterrupt2016}
\bibfield{author}{\bibinfo{person}{Jaeil Han} {and} \bibinfo{person}{Young~Man
  Kim}.} \bibinfo{year}{2016-10}\natexlab{}.
\newblock \showarticletitle{Interval-Based Adaptive Interrupt Coalescing in
  High-Speed Networks}. In \bibinfo{booktitle}{\emph{2016 {{International
  Conference}} on {{Information}} and {{Communication Technology Convergence}}
  ({{ICTC}})}} ({Jeju}). \bibinfo{publisher}{{IEEE}}, \bibinfo{pages}{68--70}.
\newblock
\showISBNx{978-1-5090-1325-8}
\urldef\tempurl%
\url{https://doi.org/10.1109/ICTC.2016.7763437}
\showDOI{\tempurl}


\bibitem[\protect\citeauthoryear{Handte, Foell, Wagner, Kortuem, and
  Marron}{Handte et~al\mbox{.}}{6 10}]%
        {handteInternetofThingsEnabledConnected2016}
\bibfield{author}{\bibinfo{person}{Marcus Handte}, \bibinfo{person}{Stefan
  Foell}, \bibinfo{person}{Stephan Wagner}, \bibinfo{person}{Gerd Kortuem},
  {and} \bibinfo{person}{Pedro~Jose Marron}.}
  \bibinfo{year}{2016-10}\natexlab{}.
\newblock \showarticletitle{An {{Internet}}-of-{{Things Enabled Connected
  Navigation System}} for {{Urban Bus Riders}}}.
\newblock  \bibinfo{volume}{3}, \bibinfo{number}{5} (\bibinfo{year}{2016-10}),
  \bibinfo{pages}{735--744}.
\newblock
\showISSN{2327-4662}
\urldef\tempurl%
\url{https://doi.org/10.1109/JIOT.2016.2554146}
\showDOI{\tempurl}


\bibitem[\protect\citeauthoryear{Hou, Kong, Zhou, Wang, Cao, and Fukud}{Hou
  et~al\mbox{.}}{8 07}]%
        {houAnalysisInterruptBehavior2018}
\bibfield{author}{\bibinfo{person}{Gang Hou}, \bibinfo{person}{Weiqiang Kong},
  \bibinfo{person}{Kuanjiu Zhou}, \bibinfo{person}{Jie Wang},
  \bibinfo{person}{Xun Cao}, {and} \bibinfo{person}{Akira Fukud}.}
  \bibinfo{year}{2018-07}\natexlab{}.
\newblock \showarticletitle{Analysis of {{Interrupt Behavior Based}} on
  {{Probabilistic Model Checking}}}. In \bibinfo{booktitle}{\emph{2018 7th
  {{International Congress}} on {{Advanced Applied Informatics}}
  ({{IIAI}}-{{AAI}})}} ({Yonago, Japan}). \bibinfo{publisher}{{IEEE}},
  \bibinfo{pages}{86--91}.
\newblock
\showISBNx{978-1-5386-7447-5}
\urldef\tempurl%
\url{https://doi.org/10.1109/IIAI-AAI.2018.00026}
\showDOI{\tempurl}


\bibitem[\protect\citeauthoryear{Murphy}{Murphy}{2012}]%
        {Murphy2012}
\bibfield{author}{\bibinfo{person}{Kevin~Patrick Murphy}.}
  \bibinfo{year}{2012}\natexlab{}.
\newblock \bibinfo{booktitle}{\emph{Machine Learning: {{A}} Probabilistic
  Perspective}}.
\newblock \bibinfo{publisher}{{MIT Press}}.
\newblock


\bibitem[\protect\citeauthoryear{Nakashima, Kusakabe, Taniguchi, and
  Amamiya}{Nakashima et~al\mbox{.}}{2002}]%
        {nakashimaDesignImplementationInterrupt2002}
\bibfield{author}{\bibinfo{person}{K. Nakashima}, \bibinfo{person}{S.
  Kusakabe}, \bibinfo{person}{H. Taniguchi}, {and} \bibinfo{person}{M.
  Amamiya}.} \bibinfo{year}{2002}\natexlab{}.
\newblock \showarticletitle{Design and Implementation of Interrupt Packaging
  Mechanism}. In \bibinfo{booktitle}{\emph{International {{Workshop}} on
  {{Innovative Architecture}} for {{Future Generation High}}-{{Performance
  Processors}} and {{Systems}}}} ({Big Island, HI, USA}).
  \bibinfo{publisher}{{IEEE Comput. Soc}}, \bibinfo{pages}{95--102}.
\newblock
\showISBNx{978-0-7695-1635-6}
\urldef\tempurl%
\url{https://doi.org/10.1109/IWIA.2002.1035023}
\showDOI{\tempurl}


\bibitem[\protect\citeauthoryear{Rajaratnam and Takawira}{Rajaratnam and
  Takawira}{1996}]%
        {rajaratnamNetworkModellingCircuitswitched1996}
\bibfield{author}{\bibinfo{person}{M. Rajaratnam} {and} \bibinfo{person}{F.
  Takawira}.} \bibinfo{year}{1996}\natexlab{}.
\newblock \showarticletitle{Network Modelling in Circuit-Switched Networks
  Using the Double Interrupted {{Poisson}} Process Model}. In
  \bibinfo{booktitle}{\emph{Proceedings of 8th {{Mediterranean Electrotechnical
  Conference}} on {{Industrial Applications}} in {{Power Systems}}, {{Computer
  Science}} and {{Telecommunications}} ({{MELECON}} 96)}} ({Bari, Italy}),
  Vol.~\bibinfo{volume}{2}. \bibinfo{publisher}{{IEEE}},
  \bibinfo{pages}{971--975}.
\newblock
\showISBNx{978-0-7803-3109-9}
\urldef\tempurl%
\url{https://doi.org/10.1109/MELCON.1996.551371}
\showDOI{\tempurl}


\bibitem[\protect\citeauthoryear{Salah}{Salah}{7 04}]%
        {salahCoalesceNotCoalesce2007}
\bibfield{author}{\bibinfo{person}{Khaled Salah}.}
  \bibinfo{year}{2007-04}\natexlab{}.
\newblock \showarticletitle{To Coalesce or Not to Coalesce}.
\newblock  \bibinfo{volume}{61}, \bibinfo{number}{4} (\bibinfo{year}{2007-04}),
  \bibinfo{pages}{215--225}.
\newblock
\showISSN{14348411}
\urldef\tempurl%
\url{https://doi.org/10.1016/j.aeue.2006.04.007}
\showDOI{\tempurl}


\bibitem[\protect\citeauthoryear{Stankovic}{Stankovic}{1994}]%
        {stankovicRealTimeOperatingSystems}
\bibfield{author}{\bibinfo{person}{John~A. Stankovic}.}
  \bibinfo{year}{1994}\natexlab{}.
\newblock \showarticletitle{Real-Time Operating Systems}. In
  \bibinfo{booktitle}{\emph{Real Time Computing}},
  \bibfield{editor}{\bibinfo{person}{Wolfgang~A. Halang} {and}
  \bibinfo{person}{Alexander~D. Stoyenko}} (Eds.). \bibinfo{publisher}{Springer
  Berlin Heidelberg}, \bibinfo{address}{Berlin, Heidelberg},
  \bibinfo{pages}{65--82}.
\newblock
\showISBNx{978-3-642-88049-0}


\bibitem[\protect\citeauthoryear{Wollschlaeger, Sauter, and
  Jasperneite}{Wollschlaeger et~al\mbox{.}}{7 03}]%
        {wollschlaegerFutureIndustrialCommunication2017}
\bibfield{author}{\bibinfo{person}{Martin Wollschlaeger},
  \bibinfo{person}{Thilo Sauter}, {and} \bibinfo{person}{Juergen Jasperneite}.}
  \bibinfo{year}{2017-03}\natexlab{}.
\newblock \showarticletitle{The {{Future}} of {{Industrial Communication}}:
  {{Automation Networks}} in the {{Era}} of the {{Internet}} of {{Things}} and
  {{Industry}} 4.0}.
\newblock  \bibinfo{volume}{11}, \bibinfo{number}{1} (\bibinfo{year}{2017-03}),
  \bibinfo{pages}{17--27}.
\newblock
\showISSN{1932-4529}
\urldef\tempurl%
\url{https://doi.org/10.1109/MIE.2017.2649104}
\showDOI{\tempurl}


\end{thebibliography}

\end{document}